\documentclass[12pt]{article}
\usepackage{amsmath,amssymb}
\def\today{\ifcase\month\or
  January\or February\or March\or April\or May\or June\or
  July\or August\or September\or October\or November\or December\fi
        \space \number\year}
\def\versionno{M.~Postma}
\catcode`\@=11
{\count255=\time\divide\count255 by 60
\xdef\hourmin{\number\count255}
\multiply\count255 by-60\advance\count255 by\time
\xdef\hourmin{\hourmin:\ifnum\count255<10 0\fi\the\count255}}
\def\ps@draft{\let\@mkboth\@gobbletwo
    \def\@oddhead{}
    \def\@oddfoot{\hbox to 7 cm{\tiny \versionno
       \hfil}\hskip -7cm\hfil\rm\thepage \hfil {\tiny\draftdate}}
    \def\@evenhead{}\let\@evenfoot\@oddfoot}
\def\draftdate{\number\day.\number\month.\number\year\ \ \ \hourmin}

\catcode`\@=12

\input epsf
\def\be{\begin{equation}}
\def\ee{\end{equation}}
\def\bea{\begin{eqnarray}}
\def\eea{\end{eqnarray}}

\renewcommand\({\left(}
\renewcommand\){\right)}

\newcommand{\dd}{{\rm d}}
\newcommand{\e}{{\rm e}}

\newcommand\mpl{M_{\rm pl}}

\newcommand{\Lag}{{\cal L}}
\newcommand{\sm}{{\sigma^\mu}}
\newcommand{\la}{\lambda}
\newcommand{\lab}{\bar{\lambda}}
\newcommand{\psb}{\bar{\psi}}

\newcommand{\dmu}{{\partial_\mu}}

\begin{document}

%\draft
%\thisfile{c}

\vskip 1cm
\begin{center}
{\Large \bf 
Chiral cosmic strings in supergravity
\vskip 0.2cm}

\vspace*{7mm}
{\ Rachel Jeannerot$^{a}$ and Marieke Postma$^{b}$}
\vspace*{.25cm}

${}^{a)}${\it Instituut-Lorentz for Theoretical Physics,
Niels Bohrweg 2, 2333 CA Leiden, The Netherlands}\\
\vspace*{.1cm} 
${}^{b)}${\it NIKHEF, Kruislaan 409, 1098 SJ Amsterdam,
The Netherlands}

%\vspace*{.3cm}

\end{center}

\vspace{0.2cm}
\begin{center}
  {\bf Abstract} 
\end{center}
\vspace{0.2cm} 
We consider F and D-term cosmic strings formed in supersymmetric
theories. Supersymmetry is broken inside the string core, but restored
outside. In global SUSY, this implies the existence of goldstino zero
modes, and the string potentially carries fermionic currents.  We show
that these zero modes do not survive the coupling to gravity, due to
the super Higgs mechanism.  Therefore the superconductivity and
chirality properties are different in global and local supersymmetry.
For example, a string formed at the end of D-term inflation is chiral
in supergravity but non-chiral in global SUSY.

\newpage

%%%%%%%%%%%%%%%%%%%%%%%%%%%%%%%%%%%%%%%%%%%%%%%%%%%%%%%%%%%%%%%%%%%%%%%%%%
%%%%%%%%%%%%%%%%%%%%%%%%%%%%%%  section 1 %%%%%%%%%%%%%%%%%%%%%%%%%%%%%%%%
%%%%%%%%%%%%%%%%%%%%%%%%%%%%%%%%%%%%%%%%%%%%%%%%%%%%%%%%%%%%%%%%%%%%%%%%%%

\section{Introduction}

It is generically believed that the early universe went through a
series of phase transitions associated with spontaneous symmetry
breaking. This implies the existence of topological defects which
form according to the Kibble mechanism \cite{Kibble}. Whereas
monopoles and domain walls are cosmologically catastrophic, cosmic
strings may have interesting cosmological properties \cite{ShelVil}.

An example of a phase transition occurring in the early universe is
the phase transition triggering the end of hybrid inflation.  Hybrid
inflation can be embedded in grand unified theories (GUTs)
\cite{hybrid}. In supersymmetric GUTs with standard hybrid inflation,
all symmetry breaking patterns which are consistent with observations
lead to the formation of cosmic strings at the end of
inflation~\cite{infl+str} (But see \cite{shifted}). Both strings and
inflation produce primordial density perturbations.  The string
contribution is constrained by the cosmic microwave background (CMB)
data, which therefore puts strong constraints on models of hybrid
inflation~\cite{CMB_constr}.\footnote{We note that predictions are
always made based on numerical simulations of Nambu-Goto strings.
They should be interpreted with care since reliable predictions can
only be made using field theory simulations. Also remember that the
evolution properties of supersymmetric string and of current carrying
strings may actually differ from standard network simulations.}

Fermions coupling to the string forming Higgs field may be confined to
the string~\cite{jackiw}.  The (massless ) zero mode excitations give
rise to currents along the string~\cite{witten}.  Of particular
interest are ``chiral strings'', strings which carry currents
travelling in one direction only: the absence of current-current
scattering enhances the stability of the chiral current. Not only do
fermionic currents influence the evolution of the string network, and
therefore alter the constraints from CMB data, they can also stabilise
string loops against gravitational collapse~\cite{vortons}.  The
requirement that these stable loops, called vortons, do not overclose
the universe also constrains the underlying physics.

Fermionic zero modes have been studied in the context of global
supersymmetry (SUSY)~\cite{N=1,nonabelian,N=2}. The number of zero
modes are given by index theorems~\cite{index}, just as in the
non-supersymmetric case.  The new feature of SUSY theories is that
SUSY is (partly) broken inside the string core.  As a result a subset
of the fermionic zero modes can be found by simply applying a
supersymmetry transformation on the bosonic string background.  In
this paper we note that these zero modes are nothing but the
goldstinos of broken supersymmetry.  It is expected that when the
theory is coupled to gravity these zero modes disappear, as the
goldstino gets eaten in the super Higgs effect.  As we will show, this
is indeed the case.  The result is that the zero mode spectrum is
completely different, and therefore the properties of the string
network, in global supersymmetry (SUSY) and in supergravity (SUGRA).

This paper is organised as follows.  In the next section we give a
review of the F and D strings, formed after symmetry breaking driven
by F and D-terms respectively.  SUSY is fully broken in the core of
F-strings.  There are then two zero modes, with opposite chirality,
corresponding to the goldstinos.  The D-string is a BPS solution. Only
half of the SUSY is broken in its core, hence there is only one
goldstino zero mode.  In the absence of a super potential the string
is chiral.  If the winding number of the Higgs field is larger than
unity, $n > 1$, the number of zero modes is multiplied by a factor
$n$.  We show how to find all goldstino modes in this case.

In section~\ref{s_sugra} we extend the results to SUGRA.  Because of
its BPS nature, the D-string is easiest to analyse. In particular we
can evaluate the SUSY transformation on the bosonic background
explicitly, to try to find the analogue of the goldstino zero mode.
However, in SUGRA the zero mode (the gravitino component) cannot be
confined to the string.  This is a consequence of the super Higgs
effect: the gravitino is massive inside the string and cannot be
localised.  The same is expected to happen for F-strings.  We conclude
that goldstino modes found in global SUSY are absent if the theory is
extended to SUGRA.

In section~\ref{s_inflation} we look at the implications for the zero
mode spectrum in the simplest realisations of F-term and D-term
inflation.  D-strings formed at the end of D-term inflation are
non-chiral in global SUSY, as there are two zero modes with opposite
chirality, the goldstino and one zero mode coming from the
superpotential.  In SUGRA the goldstino mode is absent, and D-strings
are chiral.  F-term strings are non-chiral in both SUSY and SUGRA.  In
the former case the only zero modes are the goldstinos, whereas in the
latter there are no zero-modes.  Coupling the Higgs to Majorana
fermions will give a non-zero chiral current; such a Majorana current
however is not expected to be persistent~\cite{majorana,destroyed}.

We conclude in section 5.

%%%%%%%%%%%%%%%%%%%%%%%%%%%%%%%%%%%%%%%%%%%%%%%%%%%%%%%%%%%%%%%%%%%%%%%%%%
%%%%%%%%%%%%%%%%%%%%%%%%%%%%%%  section 2 %%%%%%%%%%%%%%%%%%%%%%%%%%%%%%%%
%%%%%%%%%%%%%%%%%%%%%%%%%%%%%%%%%%%%%%%%%%%%%%%%%%%%%%%%%%%%%%%%%%%%%%%%%%

\section{F and D strings in global SUSY}

We consider $N=1$ supersymmetric $U(1)$ gauge theories.  F-term
inflation, and therefore F-strings, can also be embedded in
non-Abelian theories or in $N=2$ supersymmetric
theories~\cite{nonabelian,N=2}.  Although details change in that case,
our main conclusions concerning the goldstino zero modes remain the
same.

Our conventions are the following.  We use a Minkowski metric with
$\eta_{\mu \nu} = {\rm diag} (1,-1,-1,-1)$, and gamma matrices in the
chiral representation
\be
\gamma^\mu = \(
\begin{matrix} 
0 & \sigma^\mu \\
\bar{\sigma}^\mu & 0 
\end{matrix}
\)
\ee
with $\sigma^\mu = (1, \sigma^i)$, $\bar{\sigma}^\mu = (1,
-\sigma^i)$, and $\sigma^i$ the Pauli matrices.  For future
convenience we also give the gamma matrices in cylindrical
coordinates $(t,z, r,\theta)$: $\gamma^t = \gamma^0$, $\gamma^z =
\gamma^3$, and
\be
\gamma^r = \(
\begin{matrix} 
0 & \e^{-i\theta} \\
\e^{i \theta} & 0 
\end{matrix}
\),
\qquad \qquad
\gamma^\theta = \(
\begin{matrix} 
0 & -i\e^{-i\theta} \\
i\e^{i \theta} & 0 
\end{matrix}
\).
\ee
For spinors we use the conventions of Bailin \& Love~\cite{bailin}.

The $U(1)$ gauge theory contains an Abelian vector multiplet $V$, and
an anomaly free combination of chiral super fields $\Phi_i$ with
charges $q_i$.  Including a Fayet-Iliopoulos (FI) term, the Lagrangian
in component form reads
\begin{equation}
\Lag = \Lag_B + \Lag_F + \Lag_Y - U \ ,
\end{equation}
with
\begin{eqnarray}
\Lag_B &=& (D^{i}_\mu \phi_i)^\dagger (D^{i\mu} \phi_i)
                - \frac{1}{4} F^{\mu\nu}F_{\mu\nu} \ , \nonumber \\
\Lag_F &=& i\psi_i \sm D^{i\ast}_\mu \psb_i + i\la_i \sm \dmu \lab_i \ , 
\nonumber \\
\Lag_Y &=& {ig \sqrt{2}} q_i \phi_i^\dagger \psi_i \la 
        - W_{ij} \psi_i \psi_j + (\mbox{c.c.})
\ ,  \nonumber\\
   U   &=& |F_i|^2 + \frac{1}{2}D^2  \nonumber \\
       &=& |W_i|^2 
+      \frac{1}{2}(g \xi - {g} q_i \phi_i^\dagger \phi_i)^2 \ ,
\label{lag}
\end{eqnarray}
where $D^i_\mu = \dmu + ig q_i A_\mu$, $F_{\mu\nu} = \dmu A_\nu -
\partial_\nu A_\mu$, $W^{ij} = \delta^2 W/\delta \phi_i \delta \phi_j$
and $W^i = \delta W/ \delta \phi_i = - F^{*i}$. The chiral fermions
$\psi_i$ and gaugino $\lambda$ are left-handed Weyl fermions. They
transform under an infinitesimal SUSY transformation as
\bea 
\delta \lambda
&=&-i D \epsilon - \frac{1}{2} \sigma^\mu
\bar{\sigma}^\nu \epsilon F_{\mu \nu} , \nonumber \\
\delta \psi_i &=& \sqrt{2} \( \epsilon F_i + i
D_\mu \phi_i \sigma^\mu \bar{\epsilon} \).  
\label{dsusy}
\eea
%

%%%%%%%%%%%%%%%%%%%%%%%%%%%%%%%%%%%%%%%%%%%%%%%%%%%%%%%%%%%%%%%%%%%%%%%
\subsection{F-strings}
\label{s:F_strings}

Consider an Abelian theory without FI term, and a superpotential of
the form
\be
W = h \Phi_0(\Phi_+ \Phi_- - v^2)
\label{W_F}
\ee
with $U(1)$ charges $\pm 1,0$ for $\Phi_\pm,\Phi_0$.  This is the
superpotential which leads to F-term hybrid inflation.~\cite{hybrid}.
Inflation takes place when $\phi_0$ is slowly rolling down its
potential, while $\phi_\pm = 0$ are fixed.  It ends when the inflaton
drops below a critical value $\phi_0 < v$, and the $U(1)$ breaking
phase transition takes place.  Cosmic strings are formed at the end of
inflation through the Kibble mechanism~\cite{Kibble}.

The $U(1)$ breaking vacuum corresponds to $\phi_0=0$, $\phi_+ \phi_- =
v^2$ and $|\phi_+| = |\phi_-|$; this is the vacuum solution outside a
cosmic string.  In terms of Weyl spinors the Yukawa part of the
Lagrangian reads
\be
\Lag_y = - m_1 \chi_1 \xi_1 - m_2 \chi_2 \xi_2 + {\rm h.c.},
\ee
with $m_1 = 2 g v$, $m_2 =  \sqrt{2} h v$ and
\bea
\chi_1 =& \frac{1}{\sqrt{2}} \(
- \e^{-i\theta} \psi_+ +\e^{+i\theta} \psi_- \), 
\qquad &\xi_1 = \lambda, \nonumber\\
\chi_2 = &\frac{1}{\sqrt{2}} \(
 \e^{-i\theta} \psi_+ +\e^{+i\theta} \psi_- \),
\qquad &\xi_2 = \psi_0.
\eea
Here $\theta$ is the phase of the Higgs field, where we have written
$\phi_\pm = v \e^{\pm i \theta}$.  In the absence of the string the
phase $\theta$ can be absorbed in the fields, and can be set
consistently to zero.  The fermions can be paired into two Dirac spinors
\be
\Psi_i = 
\( \begin{matrix} 
\xi_{\alpha_i} \\ 
\bar{\chi}^{\dot{\alpha}}_i
\end{matrix} \)
\label{vac_mass}
\ee
with $i=1,2$ and $\chi_\alpha = \epsilon_{\alpha \beta} \chi^\beta
=\epsilon_{\alpha \beta} (\bar{\chi}^{\dot{\beta}})^*$.

Consider now a cosmic string background.  The Nielsen-Olesen solution
for an infinitely long string is~\cite{N=1,nielsen}
\bea
\phi_+ &=& \phi_-^\dagger = v \e^{i n \theta} f(r), \nonumber \\
A_\theta &=& - \frac{n}{g}  a(r),\nonumber\\
F_0  &=& h v^2(1-f(r)^2), 
\label{background}
\eea
and all other components zero.  The winding number $n$ is an integer.
The profile functions $f$ and $a$ obey the equations
\bea
f'' +\frac{f'}{r} - n^2 \frac{(1-a^2)}{r^2} f = h^2 v^2 (f^2-1)f,
\nonumber \\
a''-\frac{a'}{r} = - 4 g^2 v^2(1-a)f^2,
\label{profile}
\eea
with boundary conditions $f(0)=a(0)=0$ and $f(\infty)=a(\infty)=1$.
Supersymmetry is broken by the string background, but is restored on
length scales $r \gg v^{-1}$ where the string forming fields approach
their vacuum values.

The fermionic zero mode solutions solve the $(r,\, \theta)$ dependent
part of the fermion equations of motion; they are solutions to the
full $(t,z,r,\, \theta)$ dependent equations of motion with $E=0$.
Denote the zero mode by $\psi(r,\theta)$, which will be a linear
superposition of the fermionic fields in the theory.  For zero modes
which are eigenfunctions of $\sigma^3 \psi = \pm \psi$, solutions with
non-zero energy $E$ can be constructed:~\cite{jackiw,witten}
\be 
\Psi_n(t,z,r,\, \theta) 
= \psi_n(r,\theta) \e^{\mp iE(t + n z)}
\label{sol}
\ee
with $n =\pm$ the eigenvalues under $\sigma_3$, and the $\mp$ in the
exponent corresponding to positive/negative frequency solutions. The
$n=+$ mode is moving in the $-z$ direction along the string (left
chiral), the $n=-$ mode travels in the $=z$ direction (right chiral).
The dispersion relation is $E=k_z$, the modes travel at the speed of
light.

To find the fermionic excitations of zero energy, the zero modes, we
can also apply a SUSY transformation~\cite{N=1}. This is possible
since a SUSY transformation leaves the energy of a given configuration
unchanged.  Moreover, starting with a static bosonic configuration,
the fermionic excitations obtained in this way automatically satisfy
the equations of motion (with $E =0$).  The fermionic excitation found
by applying a SUSY transformation is just the goldstino of broken
global SUSY.  Using Eq.~(\ref{dsusy}) we find two independent zero
mode solutions:~\cite{N=1}
\bea
\lambda_\pm &=& \mp i \frac{n a'}{gr} \epsilon_\pm, \nonumber \\
(\psi_0)_\pm &=& \sqrt{2} h v^2
\(1-f^2\) \epsilon_\pm \nonumber,\\
({\psi}_+)_\pm &=&  \pm i \sqrt{2} v
\( f' \pm \frac{n}{r}(1-a)f \) \e^{i(n \mp 1)\theta} 
{\epsilon_\pm^*}, \nonumber\\
({\psi}_-)_\pm &=&  \pm i \sqrt{2} v
\( f' \mp \frac{n}{r}(1-a)f \) \e^{i(-n \mp 1)\theta} 
{\epsilon_\pm^*},
\label{F_zm}
\eea
Here the spinor $\epsilon_\pm$ obeys
\be
\sigma^3 \epsilon_\pm = \pm \epsilon_\pm.
\label{echiral}
\ee
Explicitly, $\epsilon_+ = \delta_+ {1 \choose 0}$ and $\epsilon_- =
\delta_- {0 \choose 1}$, with $\delta_\pm$ a complex number.  The plus
mode is left chiral.  It is a collective excitation of $\lambda,
\psi_0, {\psi}_+,{\psi}_-$, or in terms of the vacuum fields a super
position of the $\xi_i, {\chi}_i$ in Eq.~(\ref{vac_mass}).  Likewise
for the minus mode, which is a right chiral mode.

The zero modes correspond to the massless goldstinos (two chiralities)
of broken SUSY.  As SUSY is only broken inside the core of the string,
the goldstino wave function drops rapidly outside.  The number of zero
modes can be inferred from index theorems~\cite{jackiw,index}: For
each independent Yukawa coupling to $\phi$ there are $|n|$ fermionic
zero modes, with $n$ the winding number of the Higgs field, see
Eq.~(\ref{background}).  Likewise for each Yukawa coupling to $\phi^*$
there are $|n|$ zero modes, but with opposite chirality.  Since for
the $F$-string solution there are Yukawa couplings to both $\phi_+$
and $\phi_-$, i.e., to both $\phi$ and $\phi^*$, there are zero-modes
of both chiralities.  For $|n|=1$, the method of applying SUSY
transformations gives all zero modes, given by Eq.~(\ref{F_zm}).

\subsection{Zero modes for $n>1$}

For $|n| > 1$, the method of SUSY transformations gives only one zero
mode per vortex (anti-vortex) coupling, whereas the index theorem
tells there are $|n|$.  How to find these other solutions?  To do so
we note that the Lagrangian Eq.~(\ref{lag}) has a larger symmetry than
global SUSY.  It is invariant under a local SUSY transformation with a
transformation parameter $\zeta$ that satisfies
\be 
\gamma^\mu \partial_\mu \zeta(x)=0.
\label{dzeta}
\ee
This is because the variation of the global SUSY Lagrangian under a
transformation with a local SUSY parameter is of the form $\delta
{\mathcal L} = \partial_\mu \bar{\zeta} j^\mu = ... \partial_\mu
\bar{\zeta} \gamma^\mu ... =0$ by virtue of Eq.~(\ref{dzeta}).  Here
the ellipses denote functions of the various fields in the
theory.~\footnote
{In the Noether procedure to construct the local SUSY Lagrangian from
the global one, the variation of the global Lagrangian is cancelled by
adding a term ${\mathcal L} \ni a \bar{\psi}_\mu j^\mu$, with $a$ a
constant and $\psi_\mu$ the gravitino.  Looking at the SUGRA
Lagrangian the terms right-multiplying $\bar{\psi}_\mu$, i.e., the
Noether currents, all have a factor $\gamma^\mu$ in there which can be
brought to the left.}
There are two independent solutions to Eq.~(\ref{dzeta})
\be
\zeta_\pm = \e^{\pm i m \theta} r^m \epsilon_\pm.
\label{zeta}
\ee
with $m$ an integer and $\epsilon_\pm$ the projected spinors defined
in Eq.~(\ref{echiral}).  We can use the SUSY transformation with
$\zeta$ to find a whole tower of fermionic zero mode $(E=0)$ solutions
to the equations of motion.  Only $|n|$ of them will be normalisable.
This procedure gives all ``higher'' goldstino zero modes in accordance
with the index theorem.

This can also seen by looking at the equations of motion.  The
equations split in two independent sets, one set involving the upper
components of $\lambda,\psi_i$ (eigenfunctions of $\sigma^3$ with
positive eigenvalue), and one set involving the lower components
(eigenfunctions of $\sigma^3$ with negative eigenvalue).  The
fermionic equations of motion for the first set, derived from the
Lagrangian Eq.~(\ref{lag}), are four coupled equations:
\bea
&&\e^{-i\theta} ( \partial_r - \frac{i}{r} D_\theta) {\lambda^*} 
- g \sqrt{2} v f (\e^{i n \theta} \psi_{-} - \e^{-i n \theta} \psi_{+}) =0,
\\
&&\e^{-i\theta} ( \partial_r - \frac{i}{r} D_\theta)  {\psi}_{0}^* 
+i h v f (\e^{i n \theta} \psi_{-} + \e^{-i n \theta} \psi_{+}) =0,
\\
&&\e^{-i\theta} ( \partial_r - \frac{i}{r} D_\theta) 
{\psi}_{\pm}^* 
+ v f \e^{\mp i n \theta} (i h \psi_{0} \pm g \sqrt{2} \lambda) =0.
\label{eom}
\eea
To avoid notational cluttering we have omitted the subscript $1$ from
$\lambda_1$, $\psi_{01}$, ${\psi}_{\pm 1}$ to indicate upper spinor
components.~\footnote{The corresponding equations for the lower spinor
  components are obtained by replacing $\e^{-i\theta} (\partial_r -
  \frac{i}{r}D_\theta) \to -\e^{i\theta} (\partial_r +
  \frac{i}{r}D_\theta)$.}  The angular dependence can be removed by
the substitutions
\bea
\lambda &=& \tilde{\lambda}(r) \e^{i(l-1) \theta}, \nonumber \\
\psi_{0}  &=& \tilde{\psi}_{0}(r) \e^{i(l-1) \theta}, \nonumber \\
\psi_{\pm}  &=& \tilde{\psi}_{\pm}(r) \e^{i(\pm n-l)  \theta}.
\eea
An analysis of the asymptotic behaviour of the equations tells that
there are renormalisable solutions for $1 \leq l \leq |n|$ --- this is
the index theorem~\cite{jackiw,index}. The $l=1$ solutions
corresponds the the solution found by the SUSY transformation with
chiral parameters $\epsilon_\pm$.  It can be checked explicitly, using
the equations of motion for the profile functions Eq.~(\ref{profile}),
that the solutions Eq.~(\ref{F_zm}) satisfy the equations of motion.
The solutions for higher $l$ can be obtained from the solution for
$l=1$
\bea
\lambda^l &=& \e^{i (l-1) \theta} r^{(l-1)} \lambda^{1},\nonumber \\
\psi_0^l  &=& \e^{i (l-1) \theta} r^{(l-1)}\psi_0^1, \nonumber \\
\psi_\pm^l &=& \e^{i (l-1) \theta} r^{(l-1)}\psi_\pm^1
\label{high_modes}
\eea
with the superscript $l=1$ denoting the $l=1$ solution of
Eq.~(\ref{F_zm}).  The extra terms in the equation of motion, coming
from $\gamma^\mu \partial_\mu ( \e^{i (l-1) \theta} r^{(l-1)})$ cancel
among each other.  The zero modes of Eq.~(\ref{high_modes}) are
exactly the zero mode solutions obtained by performing a SUSY
transformation with $\zeta_+$ as given in Eq.~(\ref{zeta}) with $m = l
-1$.  Analogously, all right chiral modes, those involving the lower
spinor components, can be found by applying a SUSY transformation with
$\zeta_-$ with $m = l-1$.

%%%%%%%%%%%%%%%%%%%%%%%%%%%%%%%%%%%%%%%%%%%%%%%%%%%%%%%%%%%%%%%%%%%%%
\subsection{D-strings}
\label{s:D_strings}

Consider supersymmetry breaking with a FI term and two oppositely
charged chiral fields $\phi_\pm$ with charges $q_\pm = \pm 1$.  Both
fields are needed to cancel anomalies.  For now, we set the super
potential to zero: $W=0$. The bosonic potential is minimised for $\xi
= |\phi_+|^2 - |\phi_-|^2$.  The vacuum manifold is degenerate under
$|\phi_i|^2 \to |\phi_i|^2 + c$, with $c$ a real constant.  The
special point $|\phi_-| =0$, $|\phi_+| = \sqrt{\xi}$ corresponds to
the BPS solution, which conserves half of the supersymmetry, as we
will now show.

Take the following bosonic background configuration:
\begin{eqnarray} 
\phi_+  & = & \sqrt{\xi} e^{in \theta}f(r) \ , \nonumber \\
A_\theta & = & -\frac{1}{g} n a(r) \ , \nonumber \\
D & = &  g \xi (1-f(r)^2) ,
\label{boson_global}
\end{eqnarray}
and all other bosonic fields zero.  The fermions transform under
infinitesimal SUSY transformations as in Eq.~(\ref{dsusy}).  We are
interested in a background solution which leaves part of the SUSY
invariant.  So we require $\delta \lambda = 0$, $\delta \psi_+ = 0$.
($\delta \psi_- =0$ automatically for the background configuration
Eq.~(\ref{boson_global})).  This gives the BPS equations
\bea
F_{12}  \mp D &=& 0,\nonumber \\
\( \partial_r \mp \frac{i}{r} (\partial_\theta + i g A_\theta) \) \phi_+
&=& 0,
\label{BPS_global}
\eea
with $F_{12}= 1/r F_{r \theta} = 1/r \partial_r A_\theta$.  Here the
$\mp$ signs correspond to chiral transformations with $\epsilon_\pm$
defined in Eq.~(\ref{echiral}).  Only the lower sign, corresponding to
a SUSY transformation with $\epsilon_-$, gives (for $n>0$) a string
solution with positive energy density~\cite{dstrings}).~\footnote{All
equations are also valid for $n < 0$ if the substitution $n \to |n|$
and $\epsilon_+ \to \epsilon_-$ is made.  From now on we will
specialise to the $n>0$ case.} This then gives the BPS equations in
terms of the profile functions $f$ and $a$:
\bea
n\frac{a'}{r} &=& g^2 \xi (1-f^2).
\nonumber \\
f' &=& n \frac{(1-a)}{r}f,
\label{BPS_global2} 
\eea
The BPS solution to Eq.~(\ref{BPS_global2}) breaks half of the
supersymmetry.  The SUSY transformations in the string background are
invariant under a transformation with $\epsilon_-$, but not with
$\epsilon_+$. The fermion transformation with $\epsilon_+$ gives the
goldstino mode, which is confined to the string.  Explicitly:
\bea
\lambda(r,\theta) &=& -i 2 g \xi (1-f^2) \epsilon_+, \nonumber \\
{\psi}_+(r,\theta) &=& 
i 2\sqrt{2} \sqrt{\xi} \frac{n}{r} (1-a) f \e^{i(n-1)\theta}
\epsilon_+^*.
\label{D_zm}
\eea
This corresponds to a left moving mode.  In the absence of a
superpotential the Yukawa Lagrangian reads
\be
\Lag_y = i g \sqrt{2} \phi_+^* \psi_+ \lambda.
\ee
It follows from the index theorem that the SUSY transformation gives
all the zero modes for $|n| = 1$. The $l=1$ mode can be obtained from
a SUSY transformation with constant parameter $\epsilon_\pm$, whereas
the $l>1$ can be obtained formally by a space dependent SUSY
transformation with $\zeta_+$ given by Eq.~(\ref{zeta}).  Thus
for $W=0$ and $\phi_- =0$ the string has $|n|$ fermionic zero modes,
all moving in the same direction: the string is chiral.

As already noted before the vacuum manifold is degenerate under
simultaneous shifts in $|\phi_+|^2$ and $|\phi_-|^2$.  These
correspond to physically distinct vacua as the mass of the gauge boson
transforms $m_A^2 \to m_A^2 + c$.  Only in the special point in moduli
space $|\phi_+| = \sqrt{\xi}$, $|\psi_-| =0$ is the string solution
BPS. (It can be easily checked that $\delta \psi_- \neq 0$ if $\phi_-
\neq 0$, and SUSY is broken completely).

Although the vacuum is degenerate, the BPS string solution is the
solution of lowest energy.  Any non-BPS string will back react on the
vacuum, and the total energy, vacuum plus string, relaxes to the
lowest state which corresponds to $\phi_- =0$ and a BPS string. This
is dubbed the vacuum selection effect in the
literature~\cite{vac_sel}.  Hence, FI gauge symmetry breaking in
global SUSY leads to chiral strings in the absence of a superpotential.

%%%%%%%%%%%%%%%%%%%%%%%%%%%%%%%%%%%%%%%%%%%%%%%%%%%%%%%%%%%%%%%%%%%%%%
\subsection{Back reaction}

We can study the back reaction of the current on the bosonic fields
using the same method of SUSY transformations.  For the bosonic fields
in the vector multiplet, performing a SUSY transformation with
parameter $\eta$ gives
\bea      
\delta \phi &=& \sqrt{2} \eta \psi,  \nonumber \\ 
\delta F &=& i \sqrt{2} \psi \sigma^\mu \bar{\eta}, \nonumber \\ 
\delta V^\mu &=& i ( \eta \sigma^\mu \bar{\lambda} - \lambda \sigma^\mu
\bar{\eta}), \nonumber \\ 
\delta D &=& - \partial_\mu V^\mu
\eea
Denoting the upper component of a spinor by a subscript 1, the result
of SUSY transformation with $\eta_1$ is
\bea
\delta \phi = \delta F &=& 0, \nonumber \\
\delta V^z = - \delta V^0 &=& 2 {\rm Im}(\lambda^*_1 \eta_1), \nonumber \\ 
\delta D &=& 0.
\eea
The last equality is automatically for the zero energy mode, but it
also holds for modes with non-zero momentum, due to the formal
equivalence $\partial_0 \leftrightarrow \partial_z$ for the solutions
Eq.~(\ref{sol}).  These results agree with \cite{ringeval}.

%A SUSY transformation with $\eta_2$ give bosonic results which are no
%longer solutions of the BPS equations.  Moreover, they give $F \neq
%0$, $D\neq 0$; SUSY commutes with the Hamiltonian, and thus acting on
%a $E=0$ excitation it should give back a $E=0$ excitation.  So this
%SUSY transformation is broken and cannot be applied ???

%%%%%%%%%%%%%%%%%%%%%%%%%%%%%%%%%%%%%%%%%%%%%%%%%%%%%%%%%%%%%%%%%%%%%%%%%%
%%%%%%%%%%%%%%%%%%%%%%%%%%%%%%  section 3 %%%%%%%%%%%%%%%%%%%%%%%%%%%%%%%%
%%%%%%%%%%%%%%%%%%%%%%%%%%%%%%%%%%%%%%%%%%%%%%%%%%%%%%%%%%%%%%%%%%%%%%%%%%

\section{D strings in supergravity}
\label{s_sugra}

In this section we extend the analysis of fermionic zero modes for
D-strings to supergravity.  We start with a discussion of the BPS
string solution in the absence of a superpotential.  We defer a
discussion of the implications for D-term inflation, which has a
non-zero superpotential, to the next section.

%%%%%%%%%%%%%%%%%%%%%%%%%%%%%%%%%%%%%%%%%%%%%%%%%%%%%%%%%%%%%%%%%%%%%%%%%
\subsection{D-strings}

The BPS solution also exists in supergravity Ref.~(\cite{dstrings}).
The local $\tilde{U}(1)$ transformation which leaves the FI-term
invariant is a combination of the flat space gauge transformation and
a super-Weyl transformation.  It is a local R-symmetry, which induces
non-zero R-charge for the fermions and the superpotential.  We will
list here only the final results, more details and explicit
expressions can be found in Refs.~(\cite{dstrings,binetruy}).

Analogously to the global SUSY case, we consider a theory with one
chiral field $\Phi$ charged under a $\tilde{U}(1)$ (which is a local
$R$ symmetry), a minimal K\"ahler potential $K = \phi^* \phi$, minimal
gauge function $f =1$, and zero superpotential $W=0$. The bosonic
part of the Lagrangian is given by
\be
e^{-1} \Lag_b = \frac 12 \mpl^2 R + (D_\mu \phi)^\dagger (D_\mu \phi)
- \frac14 F_{\mu \nu}F^{\mu \nu} - V^D
\label{Lb_sugra}
\ee
with $V_D = \frac12 D^2$ the same as in global SUSY, see
Eq.~(\ref{lag}). The covariant derivatives on the scalar fields
contain the gauge connection $A_\mu$, just as in global SUSY.  For
vanishing $F$-terms, the fermionic Lagrangian containing only the
gauginos and chiral fermions is up to the determinant of the vierbein
of the same form as in global SUSY, but with the covariant derivatives
now containing spin, gauge, and R-symmetry connections.  Lagrangian
terms specific to supergravity are the kinetic term for the gravitino,
whose covariant derivative contains both a spin and R-connection, and
the terms mixing the gravitino with the other fermions:
\be
e^{-1} {\mathcal L}_{\rm mix}
= -\frac{i}{2} D \bar{\psi}_{\mu L} \gamma^\mu \lambda
+ (\gamma^\rho D_\rho \phi_+) 
\bar{\psi}_{\mu L} \gamma^\mu \psi_+
+ {\rm h.c.}
\label{L_mix}
\ee

The metric can be written in cylindrical coordinates
as
\be
\dd s^2 = \dd t^2 -\dd z^2 - \dd r^2 - C(r)^2 \dd \theta^2.
\label{metric}
\ee
The bosonic background for $\phi_+$ and $A_\mu$ is of the same as in
the global SUSY case, Eq.~(\ref{boson_global}), but with the profile
functions obeying different BPS equations.  The fermions transform as
in Eq.~(\ref{dsusy}). In addition there is the transformation rule for
the gravitino:
\be
\delta \psi_{\mu L} = 
2 \(\partial_\mu +\frac14 \omega_\mu^{ab} \gamma_{ab}
+\frac12 i A^B_\mu \) \epsilon,
\label{dgrav}
\ee
with $\omega_\mu^{ab}$ the spin connection, and $ A^B_\mu$ the
gravitino $\tilde{U}(1)$ connection. The condition $\delta \psi_{\mu
L} = 0$ can only be satisfied for a SUSY parameter which only depends
on $\theta$.  A globally well-behaved spinor parameter (killing
spinor) is
\be
\tilde{\epsilon}_\pm = \e^{\pm \frac12 i\theta} {\epsilon}_\pm
\label{esugra}
\ee
with ${\epsilon}_\pm$ constant spinors satisfying the chirality
condition Eq.~(\ref{echiral}).  The requirement $\delta \lambda =0,
\delta \psi =0, \psi_{\mu L} = 0$ under a SUSY transformation with
$\tilde{\epsilon}_\pm$, gives us the supergravity BPS equations
\bea
n\frac{a'}{C} &=& \mp g^2 \xi (1-f^2).
\label{BPS1} \\
f' &=& \mp n \frac{(1-a)}{C}f, 
\label{BPS2} \\
(1- C') &=& \mp A^B_\theta 
\label{BPS3}
\eea
%
%with 
%
%\be
%A^B_\theta = \frac{n}{\mpl^2} 
%\[ \xi a - \frac{C}{2 g^2} \( \frac{a'}{C} \)' \]
%\label{Uconn}
%\ee
%
As before, positivity of the string energy requires to take the lower
sign.  The solutions can be solved at the origin and at
infinity~\cite{dstrings}.  At $r \to 0$
\be
f = 0, \quad
D =g \xi, \quad
a = 0, \quad
A^B_\theta = 0, \quad
C'(0) = 1 ,
\label{sol_origin}
\ee
while at $r\to \infty$
\be
f = \sqrt{\xi}, \quad
D =0, \quad
a = 1, \quad
A^B_\theta = \frac{n\xi}{\mpl^2}, \quad
C = r(1 - \frac{n \xi}{\mpl^2}).
\label{sol_infty}
\ee
%

%%%%%%%%%%%%%%%%%%%%%%%%%%%%%%%%%%%%%%%%%%%%%%%%%%%%%%%%%%%%%%%%%%%%%
\subsection{Zero modes}
\label{s:zero_modes}

To get the zero mode, the analogue of the goldstino mode of the global
SUSY case, we proceed in the same way and perform a SUSY
transformation with $\tilde{\epsilon}_+$~\cite{Becker}.  This gives
the same solution for $\lambda$ and ${\psi}$ as in Eq.~(\ref{D_zm})
with the substitution $1/r \to 1/C(r)$, and $\epsilon \to
\tilde{\epsilon}$:
\bea
\lambda(r,\theta) &=& -i 2 g \xi (1-f^2) \e^{-i\frac12 \theta}
\epsilon_+, \nonumber \\
{\psi}_+(r,\theta) &=& 
i 2\sqrt{2} \sqrt{\xi} \frac{n}{C} (1-a) f \e^{i(n-1)\theta}
\e^{i\frac12\theta} \epsilon_+^*.
\label{sugra_zm}
\eea
In addition the $\theta$-component of the gravitino transforms
non-trivially:
\be
\psi_{\theta L} = 2 i A^B_{\theta}
= 
\left\{ 
\begin{matrix} 
0 & (r \to 0) \\
\frac{2 i n \xi}{\mpl^2} & (r \to \infty)
\end{matrix}
\right.
\label{grav_zm}
\ee
The solution does not fall off at infinity, see Eq.~(\ref{sol_infty}),
and the gravitino is not localised on the string.  The static
fermionic solution has $E=0$ since both kinetic and potential energy
vanish outside the string for a massless gravitino with constant wave
function.  However, any zero mode moving along the string with
momentum $k$ would imply infinite energy.  Likewise, the modes for $l
> 1$ lead to a gravitino wave function that does not fall off at
infinity.

It can be verified explicitly that the zero modes are solutions to the
equations of motion.  The equation of motion for the upper component
of the gaugino field for example is
\be
\e^{-i \theta} \(\partial_r - \frac{i}{C} D_\theta \) {\lambda^*}
+ g \sqrt{2 \xi} \e^{-in \theta} \psi_+ 
+ \e^{-i \theta} \frac{D}{C} {\psi}^*_{\theta L}=0.
\ee
The $\theta$ dependence cancels out for the zero mode solutions of
Eqs.~(\ref{sugra_zm},~\ref{grav_zm}).  The terms $\partial_r
{\lambda}^*$ together with the term involving $\psi_+$ cancel by the
second BPS equation (\ref{BPS2}), just as in the global SUSY case. The
covariant derivative is 
\be
D_\theta {\lambda}^* = \( \partial_\theta - \frac{i}{2} C' 
+ \frac 12 A^B_\theta \) {\lambda}^* = i A^B_\theta {\lambda}^*,
\ee
where the second equality follows from the third BPS equation
(\ref{BPS3}).  The $C'$ term comes from the spin connection
$\omega^{12}_\theta=-C'$ and $A^B_\theta$ is the R-connection.
Plugging it all in the equation of motion becomes $ A^\beta_\theta
\lambda^* + D \psi^*_{\theta L} =0$, which is indeed
identically zero for the zero mode solutions
Eqs.~(\ref{sugra_zm},~\ref{grav_zm}).

It is not surprisingly that the gravitino cannot be confined to the
string.  In the string core, where both $D$ and $\gamma^\rho D_\rho
\phi_+$ are non-zero, the fermionic fields acquire mass terms through
the couplings in $\Lag_{\rm mix}$, wee Eq.~(\ref{L_mix}) .  This is
nothing but the super Higgs effect, the goldstino gets eaten by the
gravitino which acquires a mass in the process.  Note that the
gravitino mass is maximum in the string core and zero outside. This is
the opposite from the usual fermionic fields which couple to the Higgs
field, which are massless in the core of the string, but massive
outside.  It can be understood that such fermions can get confined to
the string as their energy is lowered in the string core.  But this is
not the case for the gravitino.

We conclude that the goldstino zero mode present in the global SUSY
case has no equivalence in the SUGRA case.  Without a superpotential
the SUGRA string has no zero modes.

The SUGRA analysis has been done for the D-term string, as the
BPS-nature simplifies the equations considerably.  However, the super
Higgs effect is generic to supersymmetry breaking.  Therefore the
goldstino zero modes present in global SUSY theories are absent in the
SUGRA extension, independent of whether the symmetry is broken by D-
or F-terms, of whether the theory has $N=1$ or $N=2$ symmetry, or
whether the theory is Abelian or embedded in some non-Abelian set-up.

%%%%%%%%%%%%%%%%%%%%%%%%%%%%%%%%%%%%%%%%%%%%%%%%%%%%%%%%%%%%%%%%%%%%
%%%%%%%%%%%%%%%%%%%%%%% section 4 %%%%%%%%%%%%%%%%%%%%%%%%%%%%%%%%%%
%%%%%%%%%%%%%%%%%%%%%%%%%%%%%%%%%%%%%%%%%%%%%%%%%%%%%%%%%%%%%%%%%%%%

\section{Hybrid inflation}
\label{s_inflation}

In this section we count the number of zero modes for F- and D-strings
formed at the end of Hybrid inflation, both in global SUSY and in
SUGRA.  We limit the discussion to the standard/minimal model of
hybrid inflation.

%%%%%%%%%%%%%%%%%%%%%%%%%%%%%%%%%%%%%%%%%%%%%%%%%%%%%%%%%%%%%%%%%%%%
\subsection{D-term inflation}

We first consider the globally supersymmetric theory.  The superpotential
in standard $D$-term inflation is
\be
W = h \phi_0 \phi_+ \phi_-.
\label{D_infl}
\ee
The inflaton $\phi_0$ is neutral under $U(1)$, and the Higgs fields
$\phi_+$ and $\phi_-$ have opposite charges.
%Higher polynomials in $\phi_0$ can be forbidden by a suitably chosen
%R-symmetry.  This assures that the potential for $\phi_0$ is flat and
%slow roll inflation can take place.
The potential including the FI term is
\be
V = h^2 |\phi_0|^2 |\phi_-|^2 +  h^2 |\phi_-|^2 |\phi_+|^2
+  h^2 |\phi_+|^2 |\phi_0|^2
+ \frac{g^2}{2} (\xi -  |\phi_+|^2 +  |\phi_-|^2)^2
\label{W_DI}
\ee
It is minimised for $\phi_0 =\phi_- =0$ and $\phi_+ = \sqrt{\xi}$.
The degeneracy of the vacuum present in the absence of a super
potential is lifted.  Inflation happens for large initial inflaton
VEVs, while $\phi_0$ is slowly rolling down its potential with the
Higgses fixed at $\phi_+ = \phi_- =0$.  It ends when the inflaton
drops below the critical value $\phi_{c}= g \sqrt{\xi}/h$ and it
becomes energetically favourable for $\phi_+$ to develop a VEV.  At the
end of inflation, during the $U(1)$ breaking phase transition, cosmic
D-strings form through the Kibble mechanism.

The Yukawa terms in the Lagrangian are of the form
\be
\Lag_Y = ig\sqrt{2} \phi_+^* \psi_+ \lambda
 - h \phi_+ \psi_- \psi_0 + {\rm h.c.}
\label{yuk_DI}
\ee
Far away from the string this corresponds to two massive Dirac
particles.  Index theorems tell us that a Dirac fermion coupling to a
anti-vortex gives $|n|$ left moving zero modes proportional to
$\epsilon_+$, whereas a coupling to a vortex gives $|n|$ right moving
modes proportional to $\epsilon_-$. The left moving modes are nothing
but the goldstino modes found in section~\ref{s:D_strings} by applying
a supersymmetry transformation. The second term in $\Lag_Y$, which
comes from the superpotential, gives in addition rise to $|n|$ right
moving modes.  Therefore the D-strings formed at the end of D-term
inflation in the global supersymmetric theory are non-chiral, they
have both left and right moving modes.

D-term inflation can be generalised to supergravity,
see~\cite{binetruy} for the explicit construction.  The super
potential is the same as in global SUSY~\footnote{Since the gravitino
and gaugino carry a non-zero charge under under $\tilde{U}(1)$ the
anomaly cancellation conditions are different in SUGRA.  In particular,
three neutral fields are needed to cancel anomalies.  We will assume
that only one of them, the inflaton $\phi_0$, is coupled to the Higgs
fields $\phi_\pm$.}, and the Yukawa interactions are the same as in
the global SUSY case, Eq.~(\ref{yuk_DI}), augmented by $\Lag_{\rm
mix}$ of Eq.~(\ref{L_mix}) which includes terms involving the
gravitino.  Although details change in the SUGRA case, the index
theorem still applies.  There are thus $|n|$ right-moving modes from
the $\phi \psi_0 \psi_-$ coupling.  However, as discussed in
section~\ref{s:zero_modes}, the goldstinos are auxiliary fields, they
get eaten by the gravitino, and the goldstino modes are absent in
SUGRA.  As a result the D-strings forming at the end of D-term
inflation are chiral in the supergravity context.  There is only a
right moving zero mode, which is a superposition of $\psi_0$ and
$\psi_-$.

\subsection{$F$ term inflation} 

The strings formed at the end of F-term inflation can carry currents,
but these are non-chiral in the globally symmetric theory --- see the
discussion in section~\ref{s:F_strings}.  There are two goldstino
modes, one left and one right moving.  We expect analogously to the
D-string case, that the goldstino modes are absent in the supergravity
version of the theory, due to the super Higgs mechanism.  Therefore
the supergravity strings formed at the end of F-term inflation are not
current carrying.

The only way to get a chiral F-term string is to add a Yukawa couplings
of the form ~\cite{binetruy}.
\be
W_1 = h_1 \phi_+ \chi_{-1/2}^2 \quad {\rm or} \quad
W_2 = h_1 \frac{ \phi_+^2 \chi_{-1}^2}{ \mpl}
\label{eq:maj}
\ee
Such vortex-fermion couplings give rise to a right-moving fermion zero
modes~\cite{binetruy}.  The result is a chiral supergravity
string.

In phenomenological models, F-term inflation is often embedded in
grand unified theories.  If the symmetry broken at the end of
inflation is $U(1)_{B-L}$, $(B-L)$-cosmic strings form \cite{lept}.
The superpotential includes a coupling between the string forming
Higgs field and the right-handed Majorana neutrino is exactly of the
form Eq.~(\ref{eq:maj}).~\footnote{If $\Phi_+$ is an $SU(2)_R$
doublet, the right-handed neutrino gets its mass via a
non-normalisable term.}

Note, however, that the fermion field $\chi$ in Eq.~(\ref{eq:maj}) is
a Majorana particle, i.e., a particle which is its own
anti-particle. In this case the fermionic current cannot be protected
by lepton number conservation, and consequently it is not expected to
survive for string loops~\cite{majorana}.  Moreover, for $B-L$
strings, neutrinos zero modes are destroyed after the electroweak
phase transition \cite{destroyed}.

%%%%%%%%%%%%%%%%%%%%%%%%%%%%%%%%%%%%%%%%%%%%%%%%%%%%%%%%%%%%%%%%%%%%%%
%%%%%%%%%%%%%%%%%%%%% conclusions %%%%%%%%%%%%%%%%%%%%%%%%%%%%%%%%%%%%
%%%%%%%%%%%%%%%%%%%%%%%%%%%%%%%%%%%%%%%%%%%%%%%%%%%%%%%%%%%%%%%%%%%%%%

\section{Conclusions}
\label{s_concl}

We considered the spectrum of zero modes in both F-term and D-term
inflation.  In global SUSY (part of) the zero-modes can be obtained by
performing a SUSY transformation on the bosonic back ground.  These
zero modes are nothing than the goldstino modes, corresponding to
broken SUSY in the string core.  The goldstino zero-mode does not
survive the coupling to gravity, as it gets eaten by the gravitino in
the process --- the super Higgs effect.  As a result, the number of
zero-modes, and therefore the properties of the strings, differ in
global SUSY and SUGRA.  In particular, F-strings are current carrying
in SUSY, but not in SUGRA.  D-strings are non-chiral in SUSY, but
chiral in SUGRA.

\section*{Acknowledgements}
The authors thank Lubo Musongela and Francesco Bigazzi for useful
discussions. RJ would like to thank The Netherlands Organisation for
scientific research [NWO] for financial support.

%%%%%%%%%%%%%%%%%%%%%%%%%%%%%%%%%%%%%%%%%%%%%%%%%%%%%%%%%%%%%%%%%%%%%%%%%%
%%%%%%%%%%%%%%%%%%%%%%%%%%%  bibliography  %%%%%%%%%%%%%%%%%%%%%%%%%%%%%%%
%%%%%%%%%%%%%%%%%%%%%%%%%%%%%%%%%%%%%%%%%%%%%%%%%%%%%%%%%%%%%%%%%%%%%%%%%%


\begin{thebibliography}{99}

\bibitem{Kibble}
%\KibbleSJ
T.~W.~B.~Kibble,
%``Topology Of Cosmic Domains And Strings,''
J.\ Phys.\ A {\bf 9}, 1387 (1976).
%%CITATION = JPAGB,A9,1387;%%

\bibitem{ShelVil} 
A. Vilenkin and E.P.S. Shellard, E.P.S., 1994,
``Cosmic Strings and other Topological Defects'', (Cambridge
U. Press).
%\HindmarshRE
M.~B.~Hindmarsh and T.~W.~B.~Kibble,
%``Cosmic strings,''
Rept.\ Prog.\ Phys.\  {\bf 58}, 477 (1995)
[arXiv:hep-ph/9411342].
%%CITATION = HEP-PH 9411342;%%


\bibitem{hybrid}
%\LindeCN
A.~D.~Linde,
%``Hybrid inflation,''
Phys.\ Rev.\ D {\bf 49}, 748 (1994)
[arXiv:astro-ph/9307002].
%%CITATION = ASTRO-PH 9307002;%%
%\CopelandVG
E.~J.~Copeland, A.~R.~Liddle, D.~H.~Lyth, E.~D.~Stewart and D.~Wands,
%``False vacuum inflation with Einstein gravity,''
Phys.\ Rev.\ D {\bf 49}, 6410 (1994)
[arXiv:astro-ph/9401011].
%%CITATION = ASTRO-PH 9401011;%%
%\DvaliMS
G.~R.~Dvali, Q.~Shafi and R.~K.~Schaefer,
%``Large scale structure and supersymmetric inflation without fine tuning,''
Phys.\ Rev.\ Lett.\  {\bf 73}, 1886 (1994)
[arXiv:hep-ph/9406319].
%%CITATION = HEP-PH 9406319;%%

\bibitem{infl+str}
%\JeannerotIS
R.~Jeannerot,
%``Inflation in supersymmetric unified theories,''
Phys.\ Rev.\ D {\bf 56}, 6205 (1997)
[arXiv:hep-ph/9706391].
%%CITATION = HEP-PH 9706391;%%
%\JeannerotQV
R.~Jeannerot, J.~Rocher and M.~Sakellariadou,
%``How generic is cosmic string formation in SUSY GUTs,''
Phys.\ Rev.\ D {\bf 68}, 103514 (2003)
[arXiv:hep-ph/0308134].
%%CITATION = HEP-PH 0308134;%%

\bibitem{shifted}
%\JeannerotSV
R.~Jeannerot, S.~Khalil, G.~Lazarides and Q.~Shafi,
%``Inflation and monopoles in supersymmetric SU(4)c x SU(2)L x SU(2)R,''
JHEP {\bf 0010}, 012 (2000)
[arXiv:hep-ph/0002151].
%%CITATION = HEP-PH 0002151;%%

\bibitem{CMB_constr}
%\ContaldiQS
C.~Contaldi, M.~Hindmarsh and J.~Magueijo,
%``CMB and density fluctuations from strings plus inflation,''
Phys.\ Rev.\ Lett.\  {\bf 82}, 2034 (1999)
[arXiv:astro-ph/9809053].
%%CITATION = ASTRO-PH 9809053;%%%
%\BattyeXE
R.~A.~Battye and J.~Weller,
%``Cosmic structure formation in hybrid inflation models,''
Phys.\ Rev.\ D {\bf 61}, 043501 (2000)
[arXiv:astro-ph/9810203].
%%CITATION = ASTRO-PH 9810203;%%
N.~Bevis, M.~Hindmarsh and M.~Kunz,
%``WMAP constraints on inflationary models with global defects,''
Phys.\ Rev.\ D {\bf 70}, 043508 (2004)
[arXiv:astro-ph/0403029].
M.~Endo, M.~Kawasaki and T.~Moroi,
%``Cosmic string from D-term inflation and curvaton,''
Phys.\ Lett.\ B {\bf 569}, 73 (2003)
[arXiv:hep-ph/0304126].
%%CITATION = HEP-PH 0304126;%%
%%CITATION = ASTRO-PH 0403029;%%
%\JeongUT
E.~Jeong and G.~F.~Smoot,
%``Search for cosmic strings in CMB anisotropies,''
arXiv:astro-ph/0406432.
%%CITATION = ASTRO-PH 0406432;%%
J.~Rocher and M.~Sakellariadou,
%``The true story about the cosmological role of cosmic strings,''
arXiv:hep-ph/0405133;
%%CITATION = HEP-PH 0405133;%%
L.~Pogosian, M.~C.~Wyman and I.~Wasserman,
%``Observational constraints on cosmic strings: Bayesian analysis in a three
%dimensional parameter space,''
arXiv:astro-ph/0403268;
%%CITATION = ASTRO-PH 0403268;%%
L.~Pogosian, S.~H.~H.~Tye, I.~Wasserman and M.~Wyman,
%``Observational constraints on cosmic string production during brane
%inflation,''
Phys.\ Rev.\ D {\bf 68} (2003) 023506
[arXiv:hep-th/0304188].
%%CITATION = HEP-TH 0304188;%%



%\cite{Jackiw:1981ee}
\bibitem{jackiw}
R.~Jackiw and P.~Rossi,
%``Zero Modes Of The Vortex - Fermion System,''
Nucl.\ Phys.\ B {\bf 190} (1981) 681.
%%CITATION = NUPHA,B190,681;%%

%\cite{Witten:eb}
\bibitem{witten}
E.~Witten,
%``Superconducting Strings,''
Nucl.\ Phys.\ B {\bf 249} (1985) 557.
%%CITATION = NUPHA,B249,557;%%

\bibitem{vortons}
%\DavisIJ
R.~L.~Davis and E.~P.~S.~Shellard,
%``Cosmic Vortons,''
Nucl.\ Phys.\ B {\bf 323}, 209 (1989).
%%CITATION = NUPHA,B323,209;%%

\bibitem{N=1}
S.~C.~Davis, A.~C.~Davis and M.~Trodden,
%``N = 1 supersymmetric cosmic strings,''
Phys.\ Lett.\ B {\bf 405}, 257 (1997)
[arXiv:hep-ph/9702360].
%%CITATION = HEP-PH 9702360;%%


\bibitem{nonabelian}
S.~C.~Davis, A.~C.~Davis and M.~Trodden,
%``Cosmic strings, zero modes and SUSY breaking in nonabelian N = 1 gauge
%theories,''
Phys.\ Rev.\ D {\bf 57} (1998) 5184
[arXiv:hep-ph/9711313].
%%CITATION = HEP-PH 9711313;%%

\bibitem{N=2}
A.~Achucarro, A.~C.~Davis, M.~Pickles and J.~Urrestilla,
%``Fermion zero modes in N = 2 supervortices,''
Phys.\ Rev.\ D {\bf 68} (2003) 065006
[arXiv:hep-th/0212125];
%%CITATION = HEP-TH 0212125;%%
J.~D.~Edelstein, C.~Nunez and F.~Schaposnik,
%``Supersymmetry and Bogomolny equations in the Abelian Higgs model,''
Phys.\ Lett.\ B {\bf 329} (1994) 39
[arXiv:hep-th/9311055].
%%CITATION = HEP-TH 9311055;%%

\bibitem{index} 
E.~J.~Weinberg,
%``Index Calculations For The Fermion - Vortex System,''
Phys.\ Rev.\ D {\bf 24} (1981) 2669.
%%CITATION = PHRVA,D24,2669;%%

\bibitem{bailin} 
D. Bailin and A. Love, ``Supersymmetric Gauge Field
Theory and String Theory'' (Graduate Student Series in Physics), IOP
Publishing, 1994.


\bibitem{nielsen}
%\NielsenCS
H.~B.~Nielsen and P.~Olesen,
%``Vortex-Line Models For Dual Strings,''
Nucl.\ Phys.\ B {\bf 61}, 45 (1973).
%%CITATION = NUPHA,B61,45;%%

%\cite{Dvali:2003zh}
\bibitem{dstrings}
G.~Dvali, R.~Kallosh and A.~Van Proeyen,
%``D-term strings,''
JHEP {\bf 0401} (2004) 035
[arXiv:hep-th/0312005];
J.~D.~Edelstein, C.~Nunez and F.~A.~Schaposnik,
%``Supergravity and a Bogomolny bound in three-dimensions,''
Nucl.\ Phys.\ B {\bf 458} (1996) 165
[arXiv:hep-th/9506147];
%%CITATION = HEP-TH 9506147;%%
J.~D.~Edelstein, C.~Nunez and F.~A.~Schaposnik,
%``Bogomol'nyi Bounds and Killing Spinors in d=3 Supergravity,''
Phys.\ Lett.\ B {\bf 375} (1996) 163
[arXiv:hep-th/9512117].
%%CITATION = HEP-TH 9512117;%%

\bibitem{vac_sel}
A.~A.~Penin, V.~A.~Rubakov, P.~G.~Tinyakov and S.~V.~Troitsky,
%``What becomes of vortices in theories with flat directions,''
Phys.\ Lett.\ B {\bf 389} (1996) 13
[arXiv:hep-ph/9609257];
%%CITATION = HEP-PH 9609257;%%
A.~Achucarro, A.~C.~Davis, M.~Pickles and J.~Urrestilla,
%``Vortices in theories with flat directions,''
Phys.\ Rev.\ D {\bf 66} (2002) 105013
[arXiv:hep-th/0109097].
%%CITATION = HEP-TH 0109097;%%

%\cite{Ringeval:2000kz}
\bibitem{ringeval}
C.~Ringeval,
%``Equation of state of cosmic strings with fermionic current-carriers,''
Phys.\ Rev.\ D {\bf 63} (2001) 063508
[arXiv:hep-ph/0007015].
%%CITATION = HEP-PH 0007015;%%

%\cite{Binetruy:2004hh}
\bibitem{binetruy}
P.~Binetruy, G.~Dvali, R.~Kallosh and A.~Van Proeyen,
%``Fayet-Iliopoulos terms in supergravity and cosmology,''
arXiv:hep-th/0402046.
%%CITATION = HEP-TH 0402046;%%


\bibitem{Becker}
K.~Becker, M.~Becker and A.~Strominger,
%``Five-branes, membranes and nonperturbative string theory,''
Nucl.\ Phys.\ B {\bf 456} (1995) 130
[arXiv:hep-th/9507158].
%%CITATION = HEP-TH 9507158;%%

\bibitem{lept}
R.~Jeannerot,
%``A new mechanism for leptogenesis,''
Phys.\ Rev.\ Lett.\  {\bf 77}, 3292 (1996)
[arXiv:hep-ph/9609442].
%%CITATION = HEP-PH 9609442;%%


\bibitem{majorana}
R.~Jeannerot and M.~Postma,
%``Majorana zero modes,''
JHEP {\bf 0412} (2004) 032
[arXiv:hep-ph/0411259].
%%CITATION = HEP-PH 0411259;%%


\bibitem{destroyed}
%\DavisBU
S.~C.~Davis, A.~C.~Davis and W.~B.~Perkins,
%``Cosmic string zero modes and multiple phase transitions,''
Phys.\ Lett.\ B {\bf 408}, 81 (1997)
[arXiv:hep-ph/9705464].
%%CITATION = HEP-PH 9705464;%%

\end{thebibliography}
\end{document}